\documentclass[12pt]{article}
\usepackage{amsmath,amssymb}
\usepackage{bm}
\usepackage{color}
\usepackage{cite}
\usepackage{mathtools}

\usepackage{multirow}

\numberwithin{equation}{section}

\begin{document}

\title{
\begin{flushright}
\ \\*[-80pt]
\begin{minipage}{0.25\linewidth}
\normalsize
EPHOU-24-015\\
KYUSHU-HET-299 \\*[50pt]
\end{minipage}
\end{flushright}
{\Large \bf
Modular symmetry of localized modes
\\*[20pt]}}

\author{
Tatsuo Kobayashi$^{1}$,
Hajime Otsuka$^{2}$, 
Shohei Takada$^{1}$, 
and
~Hikaru Uchida$^{3}$
\\*[20pt]
\centerline{
\begin{minipage}{\linewidth}
\begin{center}
$^{1}${\it \normalsize
Department of Physics, Hokkaido University, Sapporo 060-0810, Japan} \\*[5pt]
$^{2}${\it \normalsize
Department of Physics, Kyushu University, 744 Motooka, Nishi-ku, Fukuoka 819-0395, Japan} \\*[5pt]
$^{3}${\it \normalsize
Institute for the Advancement of Graduate Education, Hokkaido University, Sapporo 060-0817, Japan} \\*[5pt]
\end{center}
\end{minipage}}
\\*[50pt]}

\date{
\centerline{\small \bf Abstract}
\begin{minipage}{0.9\linewidth}
\medskip
\medskip
\small
We study the modular symmetry of localized modes on fixed points of $T^2/\mathbb{Z}_2$ orbifold.
First, we find that the localized modes with even (odd) modular weight generally have $\Delta(6n^2)$ ($\Delta'(6n^2)$) modular flavor symmetry. 
Moreover, when we consider an additional Ansatz, the localized modes with even (odd) modular weight generally enjoy $S_3$ ($S'_4$) modular flavor symmetry, and we show the concrete wave functions of the localized modes.
\end{minipage}
}

\begin{titlepage}
\maketitle
\thispagestyle{empty}
\end{titlepage}

\newpage


\section{Introduction}
\label{Intro}

In higher-dimensional theories such as superstring theory, compactification of extra-dimensional space characterizes four-dimensional (4D) low-energy effective field theory.
Toroidal orbifold compactification is one of the interesting compactifications since we can obtain 4D chiral theory \cite{Dixon:1985jw,Dixon:1986jc}.
The orbifold has fixed points, which are characteristics of the orbifold different from smooth manifolds such as Calabi-Yau manifolds. 
In addition to bulk modes, localized modes on orbifold fixed points appear in heterotic orbifold models, and they are 
important modes to construct realistic models \cite{Choi:2006qh,Ibanez:2012zz}.

Localized magnetic fluxes on orbifold fixed points also induce chiral zero modes localized around the fixed points~\cite{
Lee:2003mc,Buchmuller:2015eya,Buchmuller:2018lkz,Abe:2020vmv}.
Recently, in Ref.~\cite{Kobayashi:2022tti}, it was found that the degree of freedom of localized magnetic flux, $\frac{\xi^F}{N}=\ell \in \mathbb{Z}$, leads to $\ell$-number of chiral localized zero modes.
The number of chiral localized zero modes can be counted by applying the index theorem via the resolutions of $T^2/\mathbb{Z}_N$ orbifold \cite{Kobayashi:2019fma}\footnote{See also for the index theorem on orbifolds Ref.~\cite{Aoki:2024rmf}.} and by explicitly deriving wave functions of zero modes.
Such localized magnetic fluxes can appear not only in heterotic orbifold models but in other theories, e.g. type II string theory.
Thus, the existence of localized modes is generic.

The modular symmetry is the geometrical symmetries of compact space.
Recently, the modular flavor symmetries have been attracting much attention, because finite modular groups include 
$S_3$, $A_4$, $S_4$, and $A_5$~\cite{deAdelhartToorop:2011re}, which have been used in the bottom-up approach of flavor model building \cite{Altarelli:2010gt,Ishimori:2010au,Hernandez:2012ra,King:2013eh,Kobayashi:2022moq}.
For example, in the bottom-up approach of flavor model building, 
a finite modular symmetry is assumed as the flavor symmetry, and 
Yukawa couplings and masses are written by modular forms \cite{Feruglio:2017spp}.
The modular forms of $S_3$ \cite{Kobayashi:2018vbk}, $A_4$ \cite{Feruglio:2017spp}, $S_4$ \cite{Penedo:2018nmg}, and $A_5$  \cite{Novichkov:2018nkm}, and their covering groups \cite{Liu:2019khw,Novichkov:2020eep,Liu:2020akv,Liu:2020msy,Wang:2020lxk} were studied.
One can construct interesting models leading to realistic quark and lepton mass matrices by use of those modular forms.
(See, Refs.~\cite{Kobayashi:2023zzc,Ding:2023htn}, for reviews on modular flavor symmetric models.)

In the top-down approach, the modular flavor symmetries of localized modes on fixed points 
were studied in heterotic orbifold models by stringy calculations \cite{Ferrara:1989qb,Lerche:1989cs,Lauer:1989ax,Lauer:1990tm}.
On the other hand, the modular flavor symmetries of bulk modes were studied in magnetized compactifications such as 
magnetized D-brane models~\cite{
Kobayashi:2017dyu,Kobayashi:2018rad,Kikuchi:2020frp,Kikuchi:2021ogn,
Kobayashi:2016ovu,Kobayashi:2018bff,Kariyazono:2019ehj,Ohki:2020bpo,Kikuchi:2020nxn,Almumin:2021fbk,Tatsuta:2021deu,Kikuchi:2022bkn,Kikuchi:2023clx,Ishiguro:2023jqb}, but not for localized modes.
Bulk modes in magnetized orbifold models \cite{Abe:2008fi,Abe:2013bca,Abe:2014noa} lead to interesting phenomenology.
Their coupling coefficients such as Yukawa couplings and higher order couplings \cite{Cremades:2004wa,Abe:2009dr} can be explicitly calculated from their wave functions on the orbifold, and quark and lepton mass matrices can be studied.
In addition, localized modes on orbifold fixed points induced by localized fluxes can appear, and they are generic aspects. 
The inclusion of such localized modes leads to a rich structure in magnetized compactifications.
Zero modes induced by localized fluxes may transform non-trivially under the modular symmetry like those in heterotic orbifold models.
It is interesting to study such modular transformation behavior of localized zero modes which is our purpose in this paper.
In this paper, we study the modular symmetry of zero modes localized around the orbifold fixed points by using their wave function profiles.
For simplicity, we consider two-dimensional (2D) toroidal orbifold which has $SL(2,\mathbb{Z})$ modular symmetry, that is, $T^2/\mathbb{Z}_2$ orbifold.

This paper is organized as follows.
In section~\ref{sec:review}, we review a 2D toroidal orbifold compactification.
We give a brief review on the construction in subsection~\ref{subsec:construction}, the modular symmetry and modular forms in subsection~\ref{subsec:modsym-modform}, and localized modes on orbifold fixed points in subsection~\ref{subsec:localizedmode}.
In section~\ref{sec:modoflocal}, we study the modular symmetry of the localized modes.
We study the general discussions in subsection~\ref{subsec:generality},  the specific wave functions of the localized modes in subsection~\ref{susec:wavefunction}, and various patterns of localized magnetic fluxes in subsection~\ref{sec:pattern}.
In section~\ref{sec:Conclusion}, we conclude this paper.
In Appendix A, we summarize useful relations of wave functions, discussed in subsection~\ref{subsec:generality}.

\section{Review of 2D toroidal orbifold compactification}
\label{sec:review}


\subsection{Construction of 2D torus and its orbifold}
\label{subsec:construction}

First, we review how to construct a 2D torus, $T^2$, and its $\mathbb{Z}_N$ twisted orbifold, $T^2/\mathbb{Z}_N$.
 $T^2$ can be constructed by identifying the lattice, $\Lambda$, on the complex plane, $\mathbb{C}$,~i.e. $T^2 \simeq \mathbb{C}/\Lambda$.
We introduce the complex coordinate of $\mathbb{C}$, $u$, and the two lattice vectors, $e_i\ (i=1,2)$, which span the lattice, $\Lambda$.
By use of them, the complex coordinate of $T^2$, $z$, and the complex structure modulus, $\tau$, can be written as $z=u/e_1$ and $\tau=e_2/e_1$, respectively.
Hence, the lattice identification can be written as $z+m+n\tau \sim z\ (\forall m,n \in \mathbb{Z})$.

The $T^2/\mathbb{Z}_N$ twisted orbifold can be constructed by further identifying the $\mathbb{Z}_N$ twisted points,~i.e. $e^{2\pi ik/N}z \sim z\ (\forall k \in \mathbb{Z}/N\mathbb{Z})$.
Then the fixed point $z=z_{\rm f.p.}$ on $T^2/\mathbb{Z}_N$ is defined as the point, which 
satisfies the relation $e^{2\pi ik/N}z_{\rm f.p.} + u+v\tau=z_{\rm f.p.}\ (\exists u, v \in \mathbb{Z})$.
Moreover, due to the analysis of the crystallography~\cite{Choi:2006qh}, $\tau$ should be $\tau = e^{2\pi i/N}$ for $N=3,4,6$, while $\tau$ is not constrained for $N=2$.


\subsection{Modular symmetry and Modular forms}
\label{subsec:modsym-modform}

Next, we review the modular symmetry of $T^2$ and $T^2/\mathbb{Z}_N$, and modular forms.
(See  Refs.~\cite{
Gunning:1962,Schoeneberg:1974,Koblitz:1984,Bruinier:2008,Liu:2019khw} for details of modular forms.)
Let us see a geometrical symmetry of $T^2$.
Since  the lattice vectors transformed by the following $SL(2,\mathbb{Z})$ transformation:
\begin{align}
\begin{pmatrix}
e'_2 \\ e'_1
\end{pmatrix}
=
\begin{pmatrix}
a & b \\
c & d
\end{pmatrix}
\begin{pmatrix}
e_2 \\ e_1
\end{pmatrix},
\quad
\gamma =
\begin{pmatrix}
a & b \\
c & d
\end{pmatrix}
\in SL(2,\mathbb{Z}),
\label{eq:SL2Z}
\end{align}
span  the same lattice, $T^2$ compactification has $SL(2,\mathbb{Z})$ symmetry.
Here, $SL(2,\mathbb{Z})$ is generated by the following two generators:
\begin{align}
S =
\begin{pmatrix}
0 & 1 \\
-1 & 0
\end{pmatrix},
\quad
T =
\begin{pmatrix}
1 & 1 \\
0 & 1
\end{pmatrix},
\label{eq:SL2Zgen}
\end{align}
and they satisfy the following algebraic relations:
\begin{align}
S^2 = -\mathbb{I}, \quad S^4 = (ST)^3 = \mathbb{I}. \label{eq:STrel}
\end{align}
Under the $SL(2,\mathbb{Z})$ transformation, the coordinate and the modulus, $(z,\tau)$, are transformed as
\begin{align}
\gamma: \left(z,\tau\right) \rightarrow \left( \gamma \left(z,\tau \right) \right) = \left(\frac{z}{c\tau+d},\frac{a\tau+b}{c\tau+d}\right). \label{eq:ztaumod}
\end{align}
We call it the ``modular transformation" and $SL(2,\mathbb{Z}) \equiv \Gamma$ is called the modular group.
Usually, the transformation for the modulus
is called the (inhomogeneous) modular transformation, and $\bar{\Gamma} \equiv \Gamma/\{ \pm \mathbb{I} \}$ is called (inhomogeneous) modular group. Note that $\tau$ is invariant under $S^2=-\mathbb{I}$ transformation.
We note that this modular symmetry fully remains in only $T^2/\mathbb{Z}_2$ orbifold.

Here, we also give a brief review of the ``modular forms".
We introduce the principal congruence subgroup of level $N$,
defined by
\begin{align}
\Gamma (N) \equiv \left\{ h=
\begin{pmatrix}
a' & b' \\
c' & d'
\end{pmatrix}
\in \Gamma \left|
\begin{pmatrix}
a' & b' \\
c' & d'
\end{pmatrix}
\equiv
\begin{pmatrix}
1 & 0 \\
0 & 1
\end{pmatrix}
\right.
({\rm mod}\ N)
\right\}.
\label{eq:Gamma(N)}
\end{align}
Similarly, we introduce $\bar{\Gamma}(N) \equiv \Gamma(N)/\{ \pm \mathbb{I} \}$ for $N=1,2$ and $\bar{\Gamma}(N) \equiv \Gamma(N)$ for $N>2$.
Suppose that the holomorphic function $f(z,\tau)$ transforms under the modular transformation in Eq.~(\ref{eq:ztaumod}) as
\begin{align}
&f(\gamma(z,\tau)) = J_k(\gamma, \tau) \rho(\gamma) f(z,\tau), \quad \gamma \in \Gamma, \label{eq:modform} \\
&J_k(\gamma, \tau) = (c\tau+d)^k, \label{eq:autfact} \\
&\rho(h)=\mathbb{I}, \quad h \in \Gamma(N), \ \Leftrightarrow \ \rho(S)^4=[\rho(S)\rho(T)]^3=\rho(T)^N=\mathbb{I}. \label{eq:rho}
\end{align}
Then, we call it the ``modular forms" of the integral weight $k$ for $\Gamma(N)$.
Here, $J_k(\gamma,\tau)$ is the automorphy factor with weight $k$, and $\rho(\gamma)$ is the unitary representation of the quotient group, $\Gamma'_N \equiv \Gamma/\Gamma(N)$.
In general, the modular forms are holomorphic functions of $\tau$, $f(\tau)$, which satisfy Eq.~(\ref{eq:modform}). Since $\tau$ is invariant under $S^2=-\mathbb{I}$ transformation, it should also be satisfied that $(-1)^k \rho(S)^2 = \mathbb{I}$. Hence, when $k$ is even, in particular, $\rho(\gamma)$ becomes the unitary representation of the quotient group, $\Gamma_N \equiv \bar{\Gamma}/\bar{\Gamma}(N)$. 
In particular, $\Gamma_N$ with $N=2$, $3$, $4$, and $5$ are isomorphic to $S_3$, $A_4$, $S_4$, and $A_5$, respectively~\cite{deAdelhartToorop:2011re}.
Also, $\Gamma'_N$ is the double covering group of $\Gamma_N$.


\subsection{Localized modes on orbifold fixed points}
\label{subsec:localizedmode}

Finally, we briefly review localized modes on orbifold fixed points.
Hereafter, we consider the $T^2/\mathbb{Z}_2$ orbifold since it is the only orbifold on which the modular symmetry fully remains.
The $T^2/\mathbb{Z}_2$ orbifold has four fixed points at $z=0$, $1/2$, $\tau/2$, and $(\tau+1)/2$.
Now, let us assume that localized magnetic fluxes induce $\ell$-number of chiral zero modes, $\psi_{{\rm L},z_{\rm f.p.}}^{a}(z,\tau)\ (a \in \mathbb{Z}/\ell \mathbb{Z})$, localized around each fixed point at $z=z_{\rm f.p.}$, which we call chiral localized zero modes.
Note that they are copied at the points shifted by $m+n\tau\ (\forall m,n)$.
According to Ref.~\cite{Kobayashi:2022tti}, they appear because of the localized magnetic flux, $\frac{\xi^F_{z_{\rm f.p.}}}{2}$ with the degree of freedom is $\ell$, which is inserted on each $T^2/\mathbb{Z}_2$ orbifold fixed point. 
We also assume that no bulk magnetic flux is inserted overall on $T^2/\mathbb{Z}_2$, for simplicity.
Thus, we consider the case that there are $\ell$-number of chiral localized zero modes at each fixed point and one chiral bulk zero mode.
In particular, we study the modular symmetry of the chiral localized zero modes.

\section{Modular symmetry of localized modes}
\label{sec:modoflocal}


\subsection{Modular symmetry}
\label{subsec:generality}

Now, let us study the modular symmetry of localized modes.

First, we assume that localized modes, $\psi_{{\rm L}, z_{\rm f.p.}}(z,\tau)$, behave as ``modular forms", which is verified in subsection \ref{susec:wavefunction} later. 
Second, by considering transition of fixed points under $SL(2,\mathbb{Z})$ lattice transformation, the form of $\rho(\gamma)\ (\gamma \in SL(2,\mathbb{Z}))$ can be limited. 
The $S$ transformation is the exchange between the lattice basis vectors, $e_1$ and $-e_2$.
The fixed points $1/2$ and $\tau/2$ transform each other, while the other fixed points remain.
Such behavior is clarified by use of Eq.~(\ref{eq:ztaumod}).
Under the $S$ transformation, the coordinate and the modulus are transformed as
\begin{align}
    (z',\tau')=\left(-\frac{z}{\tau},-\frac{1}{\tau} \right).
\end{align}
That leads to 
\begin{align}
    z'=z\tau'.
\end{align}
Thus, the fixed point $z=1/2$ corresponds to $z'=\tau'/2$.
Similarly, the fixed point $z=\tau/2$ corresponds to $z'=\tau\tau'/2=-1/2$, which is identified with $z=1/2$.
On the other hand, the fixed points, $z=0$ and $z=(1+\tau)/2$ correspond to 
$z'=0$ and $z'=(1+\tau')/2$, respectively. 
These are just a change of basis and coordinate.
Hence, the wave functions, $\psi_{{\rm L},1/2}$ and $\psi_{{\rm L}, \tau/2}$ exchange each other in the $S$-transformed coordinate, while 
$\psi_{{\rm L},0}$ and $\psi_{{\rm L},1/2}$ remain.
We may have ambiguity of coefficients.
Then, the $S$ transformation is represented on $(\psi_{{\rm L},0}, \psi_{{\rm L},1/2}, \psi_{{\rm L}, \tau/2}, \psi_{{\rm L}, (\tau+1)/2})^T$ as 
\begin{align}
\rho(S) =
\begin{pmatrix}
s_{11} & 0 & 0 & 0 \\
0 & 0 & s_{23} & 0 \\
0 & s_{32} & 0 & 0 \\
0 & 0 & 0 & s_{44}
\end{pmatrix}.
\label{eq:SandT-S}
\end{align}
It must be a unitary matrix.
We require $|s_{ij}|=1$.

Similarly, under the $T$ transformation, the fixed points, $z=\tau/2$ and $z=(1+\tau)/2$ transform to $z'=(1+\tau')/2$ and $z'=\tau'/2$, respectively, while 
$z=0$ and $z=1/2$ remain.
That implies that under the $T$ transformation, $\psi_{{\rm L}, \tau/2}$ and $\psi_{{\rm L}, (\tau+1)/2}$ convert into each other while $\psi_{{\rm L},0}$ and $\psi_{{\rm L},1/2}$ transform into themselves.
Thus, the $T$ transformation is represented on $(\psi_{{\rm L},0}, \psi_{{\rm L},1/2}, \psi_{{\rm L}, \tau/2}, \psi_{{\rm L}, (\tau+1)/2})^T$ as 
\begin{align}
\rho(T) =
\begin{pmatrix}
t_{11} & 0 & 0 & 0 \\
0 & t_{22} & 0 & 0 \\
0 & 0 & 0 & t_{34} \\
0 & 0 & t_{43} & 0
\end{pmatrix},
\label{eq:SandT-T}
\end{align}
where $|t_{ij}|=1$.
By the following redefinition:
\begin{align}
\psi'_{{\rm L}, 1/2} = \sqrt{\frac{s_{32}}{s_{23}}} \psi_{{\rm L}, 1/2},
\quad
\psi'_{{\rm L}, (\tau+1)/2} = \sqrt{\frac{t_{34}}{t_{43}}} \psi_{{\rm L}, (\tau+1)/2},
\label{eq:repsi}
\end{align}
the phases can be reset as $s'_{32}=s'_{23}=\sqrt{s_{23}s_{32}}$ and $t'_{43}=t'_{34}=\sqrt{t_{34}t_{43}}$.
Hereafter, we omit the prime.

Third, the localized fluxes can determine the modular weight and the phases, $s_{ij}$ and $t_{ij}$,
When no bulk magnetic flux is inserted, the bulk mode (without singular gauge transformation) is described by a constant field and the modular weight is zero.
By applying the result of Ref.~\cite{Kikuchi:2023clx}, the modular weight is shifted by the localized flux, $\ell$.
Hence, we can consider the modular weight $\ell$.
Under the following transformations:
\begin{align}
S^4 : (z,\tau) \rightarrow (e^{2\pi i}z,\tau), \label{eq:ztauS4} \\
(ST)^3 : (z,\tau) \rightarrow (e^{2\pi i}z, \tau), \label{eq:ztauST3}
\end{align}
the wave functions transform as 
\begin{align}
&\begin{pmatrix}
\psi_{{\rm L},0}(S^4(z,\tau)) \\ \psi_{{\rm L},1/2}(S^4(z,\tau)) \\ \psi_{{\rm L},\tau/2}(S^4(z,\tau)) \\ \psi_{{\rm L},(\tau+1)/2}(S^4(z,\tau))
\end{pmatrix} 
= e^{-2\pi i \ell}
\begin{pmatrix}
s_{11}^4 & 0 & 0 & 0\\
0 & s_{23}^4 & 0 & 0\\
0 & 0 & s_{23}^4 & 0\\
0 & 0 & 0 & s_{44}^4
\end{pmatrix}
\begin{pmatrix}
\psi_{{\rm L},0}(z,\tau) \\ \psi_{{\rm L},1/2}(z,\tau) \\ \psi_{{\rm L},\tau/2}(z,\tau) \\ \psi_{{\rm L},(\tau+1)/2}(z,\tau)
\end{pmatrix},
\label{eq:psiS4}\\
&\begin{pmatrix}
\psi_{{\rm L},0}((ST)^3(z,\tau)) \\ \psi_{{\rm L},1/2}((ST)^3(z,\tau)) \\ \psi_{{\rm L},\tau/2}((ST)^3(z,\tau)) \\ \psi_{{\rm L},(\tau+1)/2}((ST)^3(z,\tau))
\end{pmatrix} \notag \\
=& e^{-2\pi i \ell}
\begin{pmatrix}
s_{11}^3t_{11}^3 & 0 & 0 & 0\\
0 & s_{23}^2s_{44}t_{22}t_{34}^2 & 0 & 0\\
0 & 0 & s_{23}^2s_{44}t_{22}t_{34}^2 & 0\\
0 & 0 & 0 & s_{23}^2s_{44}t_{22}t_{34}^2
\end{pmatrix}
\begin{pmatrix}
\psi_{{\rm L},0}(z,\tau) \\ \psi_{{\rm L},1/2}(z,\tau) \\ \psi_{{\rm L},\tau/2}(z,\tau) \\ \psi_{{\rm L},(\tau+1)/2}(z,\tau)
\end{pmatrix}.
\label{eq:psiST3}
\end{align}
However, both of these relations must correspond to the boundary conditions of wave functions by $2\pi$ rotation, because both Eqs.~(\ref{eq:ztauS4}) and (\ref{eq:ztauST3}) are $2\pi$ rotation.
We consider the circle around $z=0$.
Then, the phase $e^{2\pi if}$ appears in the above boundary conditions, where $f$ corresponds to 
the total amount of the magnetic flux inserted inside the circle.\footnote{Precisely, $f$ denotes the total amount of the effective magnetic flux including the effect of the localized curvature at the orbifold fixed point. In the following analysis, we consider the effective magnetic fluxes.}
When the fixed point at $z=0$ is only contained inside the circle, only $\psi_{{\rm L},0}$ feels a total magnetic flux, $\xi^F_0=2\ell$.
Thus, the boundary conditions lead to the following relations:
\begin{align}
\begin{pmatrix}
\psi_{{\rm L},0}(S^4(z,\tau)) \\ \psi_{{\rm L},1/2}(S^4(z,\tau)) \\ \psi_{{\rm L},\tau/2}(S^4(z,\tau)) \\ \psi_{{\rm L},(\tau+1)/2}(S^4(z,\tau))
\end{pmatrix}
=
\begin{pmatrix}
e^{4\pi i \ell} & 0 & 0 & 0\\
0 & 1 & 0 & 0\\
0 & 0 & 1 & 0\\
0 & 0 & 0 & 1
\end{pmatrix}
\begin{pmatrix}
\psi_{{\rm L},0}(z,\tau) \\ \psi_{{\rm L},1/2}(z,\tau) \\ \psi_{{\rm L},\tau/2}(z,\tau) \\ \psi_{{\rm L},(\tau+1)/2}(z,\tau)
\end{pmatrix},
\label{eq:psiS4-2} \\
\begin{pmatrix}
\psi_{{\rm L},0}((ST)^3(z,\tau)) \\ \psi_{{\rm L},1/2}((ST)^3(z,\tau)) \\ \psi_{{\rm L},\tau/2}((ST)^3(z,\tau)) \\ \psi_{{\rm L},(\tau+1)/2}((ST)^3(z,\tau))
\end{pmatrix}
=
\begin{pmatrix}
e^{4\pi i \ell} & 0 & 0 & 0\\
0 & 1 & 0 & 0\\
0 & 0 & 1 & 0\\
0 & 0 & 0 & 1
\end{pmatrix}
\begin{pmatrix}
\psi_{{\rm L},0}(z,\tau) \\ \psi_{{\rm L},1/2}(z,\tau) \\ \psi_{{\rm L},\tau/2}(z,\tau) \\ \psi_{{\rm L},(\tau+1)/2}(z,\tau)
\end{pmatrix}. 
\label{eq:psiST3-2}
\end{align}
By combining the above relations, we can obtain
\begin{align}
&s_{11} = e^{3\pi i \ell/2}, \quad s_{23}=s_{44} = e^{\pi i \ell/2}, \label{eq:s11s23s44} \\
&t_{11} = e^{\pi i \ell/2}, \quad t_{22}t_{34}^2 = e^{-3\pi i \ell/2}. \label{eq:t11t23t342}
\end{align}

We note that when the fixed point at $z=1/2$ ($z=\tau/2$) is included inside the circle, the identified point at $z=-1/2$ ($z=-\tau/2$) is also included, and then $\psi_{{\rm L},1/2}$ ($\psi_{{\rm L},\tau/2}$) feels a total magnetic flux, $\xi^F_{1/2}=4\ell$.
However, we can obtain the same result in Eqs.~(\ref{eq:s11s23s44}) and (\ref{eq:t11t23t342}).
Similarly, when the fixed point at $z=(\tau+1)/2$ is included inside the circle, the identified points at $z=(\tau-1)/2$, $z=(-\tau+1)/2$, and $z=-(\tau+1)/2$ are also included, and then $\psi_{{\rm L},(\tau+1)/2}$ feels a total magnetic flux, $\xi^F_{1/2}=8\ell$.
In this case, we can arrive at the same result in Eqs.~(\ref{eq:s11s23s44}) and (\ref{eq:t11t23t342}).

Moreover, from Eqs.~(\ref{eq:SandT-S}) and (\ref{eq:SandT-T}), we can find that $N$ must be $N=2n,\ n \in \mathbb{Z}$ at least 
because of $\rho(T)^N=\mathbb{I}$.
Also,  the following relation:
\begin{align}
[ \rho(S)^{-1} \rho(T)^{-1} \rho(S) \rho(T) ]^3 = \mathbb{I}, \label{eq:SinvTinvST}
\end{align}
is satisfied.
Thus, following Refs.~\cite{deAdelhartToorop:2011re,Kikuchi:2021ogn}, Eqs.~(\ref{eq:SandT-S}) and (\ref{eq:SandT-T}) becomes the representation of $\Delta(6n^2)$ ($\Delta'(6n^2)$, which is the double covering group of $\Delta(6n^2)$) modular flavor symmetry for an even (odd) weight.

Furthermore, under the $T$ transformation, localized modes at $z=0$ and $z=1/2$ do not shift from ones at $z=0$ and $z=1/2$, respectively.
Thus, the wave functions $\psi_{{\rm L},0}$ and $\psi_{{\rm L}, 1/2}$ may change by the same phase,~i.e.,
\begin{align}
t_{22} = t_{11} = e^{\pi i \ell/2}. \label{eq:t22}
\end{align}
When we impose Eq.~(\ref{eq:t22}) as an Ansatz,\footnote{Note that this Ansatz is consistent with the analysis of wave functions in subsection \ref{susec:wavefunction}.} we can find
\begin{align}
t_{34} = e^{-\pi i \ell} = (-1)^{\ell}. \label{eq:t34}
\end{align}
However, through the redefinition, we can set
\begin{align}
t_{34} = 1 \label{eq:t34re}.
\end{align}
Therefore, the transformation matrices can be written as
\begin{align}
\rho(S) =
\begin{pmatrix}
e^{3\pi i \ell/2} & 0 & 0 & 0 \\
0 & 0 & e^{\pi i \ell/2} & 0 \\
0 & e^{\pi i \ell/2} & 0 & 0 \\
0 & 0 & 0 & e^{\pi i \ell/2}
\end{pmatrix},
\quad
\rho(T) =
\begin{pmatrix}
e^{\pi i \ell/2} & 0 & 0 & 0 \\
0 & e^{\pi i \ell/2} & 0 & 0 \\
0 & 0 & 0 & 1 \\
0 & 0 & 1 & 0
\end{pmatrix}, \label{eq:SandTre}
\end{align}
and they satisfy the following relations
\begin{align}
\begin{array}{ll}
\ \rho(S)^2 = e^{\pi i \ell} \mathbb{I}, & \\
\ [ \rho(S) \rho(T) ]^3 = \mathbb{I}, & \\
\ \rho(T)^N = \mathbb{I}, & N = \left\{
\begin{array}{ll}
2 & (\ell \in 2\mathbb{Z}) \\
4 & (\ell \in 2\mathbb{Z}+1)
\end{array}
\right., \\
\ [ \rho(S)^{-1} \rho(T)^{-1} \rho(S) \rho(T) ]^3 = \mathbb{I}, & 
\end{array}
\end{align}
which means that they are the representation of the following modular flavor symmetries:
\begin{align}
\left\{
\begin{array}{ll}
\Delta(6) \simeq S_3 & (\ell \in 2\mathbb{Z}) \\
\Delta'(24) \simeq S'_4 & (\ell \in 2\mathbb{Z}+1)
\end{array}
\right. .
\end{align}
The former group $S_3$ also appears as the modular flavor symmetry of twisted modes on fixed points 
in heterotic orbifold models \cite{Lauer:1990tm,Baur:2020jwc}.

In the following subsection, we show specific wave functions of localized modes.


\subsection{Specific wave functions of localized modes}
\label{susec:wavefunction}

In this subsection, we show specific wave functions of localized modes.
As in Ref.~\cite{Kobayashi:2022tti}, they can be obtained by modifying the wave function of the bulk mode after the singular gauge transformation.
Therefore, we first review the singular gauge transformation around an orbifold fixed point, $z=z_{{\rm f.p.}}$, and the wave function of the bulk mode after the singular gauge transformation in \ref{sususec:bulkmode}.
Then, we show specific wave function of the localized modes, and study their modular flavor symmetry in \ref{sususec:localizemode}.

\subsubsection{Wave function of the bulk mode after the singular gauge transformation}
\label{sususec:bulkmode}

To define the singular gauge transformation around an orbifold fixed point, $z=z_{{\rm f.p.}}$, we introduce the wave function, $\psi_{z_{{\rm f.p.}},\mathbb{Z}_N^1}(z,\tau)$, such that it can be approximated around $z=z_{{\rm f.p.}}$ as
\begin{align}
\psi_{z_{{\rm f.p.}},\mathbb{Z}_N^1}(z,\tau) \simeq c (z-z_{{\rm f.p.}}), \label{eq:aprox}
\end{align}
where $c$ denotes a constant.
Note that it satisfies $\psi_{z_{{\rm f.p.}},\mathbb{Z}_N^1}(z_{{\rm f.p.}},\tau)=0$.
Specifically, $\psi_{z_{{\rm f.p.}},\mathbb{Z}_N^1}(z,\tau)$ can be written as~\cite{Sakamoto:2020pev} 
\begin{align}
\begin{array}{l}
\psi_{0,\mathbb{Z}_N^1}(z,\tau) = \psi^{(1/2,1/2)}(z,\tau) \\
\psi_{1/2,\mathbb{Z}_N^1}(z,\tau) = \psi^{(1/2,0)}(z,\tau) \\
\psi_{\tau/2,\mathbb{Z}_N^1}(z,\tau) = \psi^{(0,1/2)}(z,\tau) \\
\psi_{(\tau+1)/2,\mathbb{Z}_N^1}(z,\tau) = \psi^{(0,0)}(z,\tau)
\end{array},
\label{eq:zeropoint}
\end{align}
where $\psi^{(\alpha_1,\alpha_2)}(z,\tau)$ can be expressed as \cite{Cremades:2004wa}
\begin{align}
\psi^{(\alpha_1,\alpha_2)}(z,\tau) 
&= e^{2\pi i \alpha_1 \alpha_2} e^{\pi iz\frac{{\rm Im}z}{{\rm Im}\tau}} \vartheta
\begin{bmatrix}
\alpha_1 \\ -\alpha_2
\end{bmatrix}
\left( z, \tau \right)
\notag \\
&= e^{2\pi i \alpha_1 \alpha_2} e^{\pi iz\frac{{\rm Im}z}{{\rm Im}\tau}} \sum_{n \in \mathbb{Z}} e^{\pi i \tau (n+\alpha_1)^2} e^{2\pi i(z-\alpha_2)(n+\alpha_1)}.
\label{eq:waveT2}
\end{align}
By using $\psi_{z_{{\rm f.p.}},\mathbb{Z}_N^1}(z,\tau)$, the singular gauge transformation~\cite{Kobayashi:2022tti}\footnote{See also Refs.~\cite{Lee:2003mc,Buchmuller:2015eya,Buchmuller:2018lkz}.} around $z=z_{{\rm f.p.}}$ can be written as
\begin{align}
U^{\ell N}_{z_{{\rm f.p.}}}(z)
&= \left[ \left( \psi_{z_{{\rm f.p.}},\mathbb{Z}_N^1}(z,\tau)/\overline{\psi_{z_{{\rm f.p.}},\mathbb{Z}_N^1}(z,\tau)} \right)^{1/2} \right]^{\ell N}
\simeq \left[ ( z - z_{{\rm f.p.}} ) / (\overline{z - z_{{\rm f.p.}}}) \right]^{\ell N/2} \notag \\
&= \left[ \psi_{z_{{\rm f.p.}},\mathbb{Z}_N^1}(z,\tau)/\left| \psi_{z_{{\rm f.p.}},\mathbb{Z}_N^1}(z,\tau) \right| \right]^{\ell N}
\simeq \left[ ( z - z_{{\rm f.p.}} ) / \left| z - z_{{\rm f.p.}} \right| \right]^{\ell N}. 
\label{eq:singulargauge}
\end{align}
Then, we can express the bulk mode after the singular gauge transformation around the fixed point, $z=z_{{\rm f.p.}}$, as
\begin{align}
\psi_{{\rm B}}(z,\tau) = ({\rm Im}\tau)^{-\ell N/4} \left[ \psi_{z_{{\rm f.p.}},\mathbb{Z}_N^1}(z,\tau)/\left| \psi_{z_{{\rm f.p.}},\mathbb{Z}_N^1}(z,\tau) \right| \right]^{\ell N}.
\label{eq:bulkmode}
\end{align}

\subsubsection{Wave functions of localized modes and their modular symmetry}
\label{sususec:localizemode}

Now, let us show wave functions of localized modes by modifying the wave function of the bulk mode in Eq.~(\ref{eq:bulkmode}), and study their modular flavor symmetry.
Here, we note that $\left( \psi_{z_{{\rm f.p.}},\mathbb{Z}_N^1}(z,\tau) \right)^N$ in Eq.~(\ref{eq:bulkmode}) is $\mathbb{Z}_N$ invariant although the $\mathbb{Z}_N$ charge of $\psi_{z_{{\rm f.p.}},\mathbb{Z}_N^1}(z,\tau)$ is $1$.
Thus, by introducing the $\mathbb{Z}_N$ invariant mode, $\psi_{z_{{\rm f.p.}},\mathbb{Z}_N^0}(z,\tau)$, and replacing $\left( \psi_{z_{{\rm f.p.}},\mathbb{Z}_N^1}(z,\tau) \right)^N$ with $\left( \psi_{z_{{\rm f.p.}},\mathbb{Z}_N^0}(z,\tau) \right)^N$, 
the $a$-th localized mode can be expressed as~\cite{Kobayashi:2022tti}
\begin{align}
\psi_{{\rm L}}^{a}(z,\tau) =
\left( \psi_{z_{{\rm f.p.}},\mathbb{Z}_N^0}(z,\tau)/\psi_{z_{{\rm f.p.}},\mathbb{Z}_N^1}(z,\tau) \right)^{(\ell -a) N}
\psi_{{\rm B}}(z,\tau).
\label{eq:localizedmode}
\end{align}

From now, we discuss the specific wave function of $\psi_{z_{{\rm f.p.}},\mathbb{Z}_N^0}(z,\tau)$.
First, let us consider $z_{{\rm f.p.}}=0$ case.
In this case, there are three candidates of $\psi_{0,\mathbb{Z}_N^0}(z,\tau)$,~i.e., $\psi^{(0,0)}(z,\tau)$, $\psi^{(1/2,0)}(z,\tau)$, and $\psi^{(0,1/2)}(z,\tau)$.
Thus, the localized mode at $z_{{\rm f.p.}}=0$ can be written by
\begin{align}
&\psi_{{\rm L},0}^{a}(z,\tau) = ({\rm Im}\tau)^{-\ell N/4} \left[ \psi^{(1/2,1/2)}(z,\tau)/\left| \psi^{(1/2,1/2)}(z,\tau) \right| \right]^{\ell N} \left[ \Phi_0(z,\tau) / \left( \psi^{(1/2,1/2)}(z,\tau) \right)^{N} \right]^{\ell - a}, \notag \\
&\Phi_0(z,\tau) =
C_0^{(0,0)}(\tau) \left( \psi^{(0,0)}(z,\tau) \right)^{N} + C_0^{(1/2,0)}(\tau) \left( \psi^{(1/2,0)}(z,\tau) \right)^{N} + C_0^{(0,1/2)}(\tau) \left( \psi^{(0,1/2)}(z,\tau) \right)^{N},
\label{eq:localizedmode0}
\end{align}
where $C_0^{(\alpha_1, \alpha_2)}(\tau)$ denote coefficients that depend on not $z$ but $\tau$.
In particular, in order for the localized modes to be linearly transformed under the modular transformation, we set the coefficients, $C_0^{(\alpha_1, \alpha_2)}(\tau)$, as
\begin{align}
\begin{array}{l}
C_0^{(0,0)}(\tau) = \left( \psi^{(1/2,0)}(0,\tau) \psi^{(0,1/2)}(0,\tau) \right)^{N}, \\
C_0^{(1/2,0)}(\tau) = \left( \psi^{(0,0)}(0,\tau) \psi^{(0,1/2)}(0,\tau) \right)^{N}, \\
C_0^{(0,1/2)}(\tau) = \left( \psi^{(0,0)}(0,\tau) \psi^{(1/2,0)}(0,\tau) \right)^{N}.
\end{array}
\label{eq:coefficients}
\end{align}
Here, we consider the modular transformation of
$\psi^{(\alpha_1,\alpha_2)}(z,\tau)$, summarized in Eq.~(\ref{eq:eachmodulartransf}).
Namely, the localized mode at $z_{{\rm f.p.}}=0$ can be expressed as
\begin{align}
\psi_{{\rm L},0}^{a}(z,\tau) =& ({\rm Im}\tau)^{-\ell N/4} \left[ \psi^{(1/2,1/2)}(z,\tau)/\left| \psi^{(1/2,1/2)}(z,\tau) \right| \right]^{\ell N} \times \label{eq:localizedmode0re}\\
&\left[ \left\{ \left( \psi^{(0,0)}(z,\tau) \psi^{(1/2,0)}(0,\tau) \psi^{(0,1/2)}(0,\tau) \right)^{N} \right. \right. \notag \\
&\quad + \left( \psi^{(0,0)}(0,\tau) \psi^{(1/2,0)}(z,\tau) \psi^{(0,1/2)}(0,\tau) \right)^{N} \notag \\
&\quad \left. \left. + \left( \psi^{(0,0)}(0,\tau) \psi^{(1/2,0)}(0,\tau) \psi^{(0,1/2)}(z,\tau) \right)^{N} \right\} / \left( \psi^{(1/2,1/2)}(z,\tau) \right)^{N} \right]^{\ell - a}, \notag
\end{align}
and it transforms under the $S$ and $T$ transformations as
\begin{align}
\begin{array}{ll}
\psi_{{\rm L},0}^{a}(S(z,\tau)) = (-\tau)^{2(\ell -a)+\ell} e^{3\pi i\ell/2} \psi_{{\rm L},0}^{a}(z,\tau), & \psi_{{\rm L},0}^{a}(T(z,\tau)) = e^{\pi i \ell/2} \psi_{{\rm L},0}^{a}(z,\tau).
\end{array}
\label{eq:modulartransf0}
\end{align}
Note again that since we chose the coefficients as Eq.~(\ref{eq:coefficients}), all of the three terms of $\Phi_0(z,\tau)$ in Eq.~(\ref{eq:localizedmode0}) have the same behavior under the modular transformation. 
Also, the form of the wave function (\ref{eq:localizedmode0re}) has the following permutation symmetry:
\begin{align}
    \psi^{(0,0)}(z,\tau) \to \psi^{(1/2,0)}(z,\tau) \to \psi^{(0,1/2)}(z,\tau) \to 
    \psi^{(0,0)}(z,\tau), \notag \\
    \psi^{(1/2,0)}(0,\tau) \to \psi^{(0,1/2)}(0,\tau) \to \psi^{(0,0)}(0,\tau) \to 
    \psi^{(1/2,0)}(0,\tau).
    \label{eq:permutation}
\end{align}

Similarly, by replacing $z$ with $Z_{z_{\rm f.p.} } = z - z_{\rm f.p.}$ and considering the relations in Eqs.~(\ref{eq:Z12})-(\ref{eq:Ztau12}), the localized modes at the other fixed points, $z=1/2$, $\tau/2$, and $(\tau+1)/2$, can be expressed as
\begin{align}
\psi_{{\rm L},1/2}^{a}(z,\tau) =& ({\rm Im}\tau)^{-\ell N/4} \left[ \psi^{(1/2,0)}(z,\tau)/\left| \psi^{(1/2,0)}(z,\tau) \right| \right]^{\ell N} \times \label{eq:localizedmode12}\\
&\left[ \left\{ \left( \psi^{(0,1/2)}(z,\tau) \psi^{(1/2,1/2)}(0,\tau) \psi^{(0,0)}(0,\tau) \right)^{N} \right. \right. \notag \\
&\quad + \left( \psi^{(0,1/2)}(0,\tau) \psi^{(1/2,1/2)}(z,\tau) \psi^{(0,0)}(0,\tau) \right)^{N} \notag \\
&\quad \left. \left. + \left( \psi^{(0,1/2)}(0,\tau) \psi^{(1/2,1/2)}(0,\tau) \psi^{(0,0)}(z,\tau) \right)^{N} \right\} / \left( \psi^{(1/2,0)}(z,\tau) \right)^{N} \right]^{\ell - a}, \notag
\\
\psi_{{\rm L},\tau/2}^{a}(z,\tau) =& ({\rm Im}\tau)^{-\ell N/4} \left[ \psi^{(0,1/2)}(z,\tau)/\left| \psi^{(0,1/2)}(z,\tau) \right| \right]^{\ell N} \times \label{eq:localizedtau2}\\
&\left[ \left\{ \left( \psi^{(1/2,0)}(z,\tau) \psi^{(0,0)}(0,\tau) \psi^{(1/2,1/2)}(0,\tau) \right)^{N} \right. \right. \notag \\
&\quad + \left( \psi^{(1/2,0)}(0,\tau) \psi^{(0,0)}(z,\tau) \psi^{(1/2,1/2)}(0,\tau) \right)^{N} \notag \\
&\quad \left. \left. + \left( \psi^{(1/2,0)}(0,\tau) \psi^{(0,0)}(0,\tau) \psi^{(1/2,1/2)}(z,\tau) \right)^{N} \right\} / \left( \psi^{(0,1/2)}(z,\tau) \right)^{N} \right]^{\ell - a}, \notag
\\
\psi_{{\rm L},(\tau+1)/2}^{a}(z,\tau) =& ({\rm Im}\tau)^{-\ell N/4} \left[ \psi^{(0,0)}(z,\tau)/\left| \psi^{(0,0)}(z,\tau) \right| \right]^{\ell N} \times \label{eq:localizedmodetau12}\\
&\left[ - \left\{ \left( \psi^{(1/2,1/2)}(z,\tau) \psi^{(0,1/2)}(0,\tau) \psi^{(1/2,0)}(0,\tau) \right)^{N} \right. \right. \notag \\
&\quad + \left( \psi^{(1/2,1/2)}(0,\tau) \psi^{(0,1/2)}(z,\tau) \psi^{(1/2,0)}(0,\tau) \right)^{N} \notag \\
&\quad \left. \left. + \left( \psi^{(1/2,1/2)}(0,\tau) \psi^{(0,1/2)}(0,\tau) \psi^{(1/2,0)}(z,\tau) \right)^{N} \right\} / \left( \psi^{(0,0)}(z,\tau) \right)^{N} \right]^{\ell - a}, \notag
\end{align}
respectively.
Note that coefficients $C_{z_{\rm f.p.}}^{(\alpha_1,\alpha_2)}=\psi^{(\alpha'_1,\alpha'_2)}(0,\tau)\psi^{(\alpha''_1,\alpha''_2)}(0,\tau)$ are also shifted by $z_{\rm f.p.}$ and then we set $z=0$.
They transform under the $S$ and $T$ transformations as
\begin{align}
\begin{array}{ll}
\psi_{{\rm L},1/2}^{a}(S(z,\tau)) = (-\tau)^{2(\ell -a)+\ell} e^{\pi i\ell/2} \psi_{{\rm L},\tau/2}^{a}(z,\tau), & \psi_{{\rm L},1/2}^{a}(T(z,\tau)) = e^{\pi i \ell/2} \psi_{{\rm L},1/2}^{a}(z,\tau),
\\
\psi_{{\rm L},\tau/2}^{a}(S(z,\tau)) = (-\tau)^{2(\ell -a)+\ell} e^{\pi i\ell/2} \psi_{{\rm L},1/2}^{a}(z,\tau), & \psi_{{\rm L},\tau/2}^{a}(T(z,\tau)) = \psi_{{\rm L},(\tau+1)/2}^{a}(z,\tau),
\\
\psi_{{\rm L},(\tau+1)/2}^{a}(S(z,\tau)) = (-\tau)^{2(\ell -a)+\ell} e^{\pi i\ell/2} \psi_{{\rm L},(\tau+1)/2}^{a}(z,\tau), & \psi_{{\rm L},(\tau+1)/2}^{a}(T(z,\tau)) = \psi_{{\rm L},\tau/2}^{a}(z,\tau).
\end{array}
\label{eq:modulartransfother}
\end{align}
Note again that the above choice of coefficients $C_{z_{\rm f.p.}}^{(\alpha_1,\alpha_2)}$ is important to lead to the modular transformation behavior, and there are permutation symmetries similar to Eq.~(\ref{eq:permutation}) .

Therefore, we conclude that the localized modes transform under the modular transformation as
\begin{align}
\begin{array}{ll}
\psi_{{\rm L},z_{\rm f.p.}}^{a}(S(z,\tau)) = (-\tau)^{2(\ell -a)+\ell} \rho(S) \psi_{{\rm L},z_{\rm f.p.}}^{a}(z,\tau), & \psi_{{\rm L},z_{\rm f.p.}}^{a}(T(z,\tau)) = \rho(T) \psi_{{\rm L},z_{\rm f.p.}}^{a}(z,\tau), \\
\\
\rho(S) =
\begin{pmatrix}
e^{3\pi i \ell/2} & 0 & 0 & 0 \\
0 & 0 & e^{\pi i \ell/2} & 0\\
0 & e^{\pi i \ell/2} & 0 & 0 \\
0 & 0 & 0 & e^{\pi i \ell/2}
\end{pmatrix}, &
\rho(T) =
\begin{pmatrix}
e^{\pi i \ell/2} & 0 & 0 & 0 \\
0 & e^{\pi i \ell/2} & 0 & 0 \\
0 & 0 & 0 & 1 \\
0 & 0 & 1 & 0 
\end{pmatrix},
\end{array}
\label{eq:modulartransflocalizedmode}
\end{align}
which correspond to Eq.~(\ref{eq:SandTre}).
Namely, the localized modes have the following modular flavor symmetry:
\begin{align}
\left\{
\begin{array}{ll}
S_3 & (\ell \in 2\mathbb{Z}) \\
S'_4 & (\ell \in 2\mathbb{Z}+1)
\end{array}
\right. ,
\label{eq:S3S4}
\end{align}
with the modular weight $2(\ell -a)+\ell$.
In more detail, for $\ell \in 2\mathbb{Z}+1$, the localized mode at $z_{\rm f.p.}=0$ behaves as a $S'_4$ singlet while the other localized modes at $z_{\rm f.p.} = 1/2$, $\tau/2$, and $(\tau+1)/2$ behave as a $S'_4$ triplet.
On the other hand, for $\ell \in 2\mathbb{Z}$, the localized mode at $z_{\rm f.p.}=0$ behaves as a $S_3$ singlet while the other localized modes at $z_{\rm f.p.} = 1/2$, $\tau/2$, and $(\tau+1)/2$ behave as $S_3$ doublet and singlet since the representations can be rewritten as
\begin{align}
\rho(S)_3 = e^{\pi i \ell/2}
\begin{pmatrix}
-1 & 0 & 0 \\
0 & 1 & 0\\
0 & 0 & 1
\end{pmatrix},
\quad
\rho(T)_3 =
\begin{pmatrix}
\pm \frac{1}{2} & 0 & -\frac{\sqrt{3}}{2} \\
0 & \pm 1 & 0\\
-\frac{\sqrt{3}}{2} & 0 & \mp \frac{1}{2}
\end{pmatrix},
\label{eq:S3doubletsinglet}
\end{align}
by the following tri-bi-maximal transformation,
\begin{align}
V =
\begin{pmatrix}
1 & 0 & 0 \\
0 & \sqrt{\frac{2}{3}} & \pm \sqrt{\frac{1}{3}} \\
0 & \mp \sqrt{\frac{1}{3}} & \sqrt{\frac{2}{3}}
\end{pmatrix}
\begin{pmatrix}
\sqrt{\frac{1}{2}} & -\sqrt{\frac{1}{2}} & 0\\
\sqrt{\frac{1}{2}} & \sqrt{\frac{1}{2}} & 0\\
0 & 0 & 1
\end{pmatrix}
=
\begin{pmatrix}
\sqrt{\frac{1}{2}} & -\sqrt{\frac{1}{2}} & 0 \\
\sqrt{\frac{1}{3}} & \sqrt{\frac{1}{3}} & \pm \sqrt{\frac{1}{3}} \\
\mp \sqrt{\frac{1}{6}} & \mp \sqrt{\frac{1}{6}} & \sqrt{\frac{2}{3}}
\end{pmatrix}.
\label{eq:tribimaximal}
\end{align}
Here, we use the upper (lower) sign for $\ell \in 4\mathbb{Z}$ ($\ell \in 2(2\mathbb{Z}+1)$).


\subsection{Different localized magnetic fluxes}
\label{sec:pattern}

So far, we have studied the case that the same sizes of magnetic fluxes 
$\xi^F_{z_{\rm f.p.}}$ are localized on all of the fixed points.
Here, we discuss the cases that  magnetic fluxes localized at fixed points
are different from each other.
Note that the localized mode at $z=0$ is a singlet.
Thus, when we set only the magnetic flux localized at $z=0$ which is different from the others and 
the others are the same, the flavor group is the same, i.e., $S_3$ or $S'_4$.
However, only the representation of localized mode at $z=0$ changes.
For example, when $\ell \in 2(2\mathbb{Z}+1)$ and $4\mathbb{Z}$ for $z=0$,  the localized mode at $z=0$ is the non-trivial and trivial singlets of $S_3$, respectively, and it is $\mathbb{Z}_2$-odd and -even mode where $\mathbb{Z}_2$ is a subgroup of $S_3$.
On the other hand, when  $\ell \in 2\mathbb{Z}+1$, the localized mode at $z=0$ has a $\mathbb{Z}_4$ charge, where $\mathbb{Z}_4$ is a subgroup of $S'_4$.

When we introduce different magnetic fluxes at three fixed points at $z=1/2, 
\tau/2$, and $z=(1+\tau)/2$, the flavor symmetries $S_3$ and $S'_4$ are broken down.
However, when the localized fluxes at $z=\tau/2$ and $z=(1+\tau)/2$ are the same, but 
different from on $z=1/2$, the $T$ symmetry remains and it is represented by 
\begin{align}
\rho(T) =
\begin{pmatrix}
e^{\pi i \ell/2} & 0 & 0 & 0\\
0 & e^{\pi i \ell'/2} & 0 & 0\\
0 & 0 & 0 & 1 \\
0 & 0 & 1 & 0 
\end{pmatrix},
\end{align}
but the $S$ symmetry is broken.
The flavor symmetry is Abelian.
Both the localized modes at $z=0$ and $1/2$ have $\mathbb{Z}_{4}$ charge for $\ell, \ell'=$ odd and 
$\mathbb{Z}_2$ charge for $\ell, \ell'=$ even.
On the other hand, linear combinations of localized modes at $z=\tau/2$ and $(1+\tau)/2$ 
correspond to $\mathbb{Z}_2$-even and odd modes.

Similarly, when the localized fluxes at $z=1/2$ and $z=\tau/2$ are the same, but 
different from one on $z=(1+\tau)/2$, the $S$ symmetry remains and it is represented by 
\begin{align}
\rho(S) =
\begin{pmatrix}
e^{3\pi i \ell/2} & 0 & 0 & 0 \\
0 & 0 & e^{\pi i \ell'/2} & 0 \\
0 & e^{\pi i \ell'/2} & 0 & 0 \\
0 & 0 & 0 & e^{\pi i \ell''/2}
\end{pmatrix}, 
\end{align}
but the $T$ symmetry is broken.
The flavor symmetry is Abelian.
Both the localized modes at $z=0$ and $(1+\tau)/2$ have $\mathbb{Z}_{4}$ charge for $\ell, \ell''=$ odd and 
$\mathbb{Z}_2$ charge for $\ell, \ell''=$ even.
On the other hand, linear combinations of localized modes at $z=1/2$ and $\tau/2$ 
correspond to $\mathbb{Z}_2$-even and odd modes.

We have discussed flavor symmetries by several patterns of localized magnetic fluxes.
Certain patterns may be constrained by some reasons.
The degree of freedom of localized fluxes is equivalent to the Fayet-Iliopoulos (FI) terms localized at fixed points \cite{Lee:2003mc}.
In Refs.~\cite{Lee:2003mc,Abe:2020vmv}, it was found that 
non-vanishing FI terms modify the profiles of bulk zero modes and massive Kaluza-Klein modes.
Modified profiles of these modes induce radiative corrections on FI terms.
Then, such radiative corrections modify profiles of bulk modes further.
Thus, this system may be unstable.
One of stable configurations is the pattern that the sizes of FI terms at all of the fixed points are the same \cite{Abe:2020vmv}.
Therefore, such a pattern of FI terms localized fluxes is favorable from the viewpoint of stability.

Another reason can be discussed by anomaly.
In general, the non-Abelian discrete flavor symmetries such as $S_3$ and $S'_4$ can be anomalous depending on massless modes \cite{Krauss:1988zc,Ibanez:1991hv,Banks:1991xj,Araki:2008ek,Chen:2015aba,Kobayashi:2021xfs}.
However, note that $\rho(S)$ and $\rho(T)$ in Eq.~(\ref{eq:modulartransflocalizedmode}) lead to 
$\det \rho(S) = \det \rho(T)=1$.
Thus, this pattern of the localized magnetic fluxes leads to the anomaly-free flavor symmetries, 
but other patterns may lead to anomalies.

Hence, the same sizes of magnetic fluxes at all of the fixed points are favorable from the viewpoints of stability of the system and anomaly.
Then, such a pattern leads to a maximal flavor symmetry $S_3$ and $S'_4$.


\section{Conclusion}
\label{sec:Conclusion}

We have studied the modular symmetry of localized modes on fixed points of $T^2/\mathbb{Z}_2$ orbifold.
First, we find that the localized modes with even (odd) modular weight generally have $\Delta(6n^2)$ ($\Delta'(6n^2)$) modular flavor symmetry, where the order of $\rho(T)$ is $2n$.
Moreover, when we consider an additional Ansatz, we find that the localized modes with even (odd) modular weight generally have $S_3$ ($S'_4$) modular flavor symmetry.
Indeed, we have shown the specific wave functions of the localized modes.
In particular, the localized mode with even (odd) modular weight at $z_{\rm f.p.}=0$ behaves as $S_3$ ($S'_4$) singlet while the other localized modes at $z_{\rm f.p.}=1/2$, $\tau/2$, and $(\tau+1)/2$ behave as $S_3$ doublet and singlet ($S'_4$ triplet).

The flavor symmetries $S_3$ and $S'_4$ can be realized when the sizes of localized fluxes at 
all of the fixed points are the same.
In general, localized fluxes at fixed points can be different from each other.
Such patterns lead to smaller flavor symmetries such as Abelian symmetries.
However, some patterns of localized fluxes may lead to unstable configurations of localized fluxes by radiative corrections.
Also those Abelian symmetries can be anomalous.
Hence, the same sizes of localized fluxes at all of the fixed points are favorable 
from these viewpoints.
They lead to the maximal flavor symmetries $S_3$ and $S'_4$.

In this paper, we assumed that no bulk magnetic flux is inserted overall on $T^2/\mathbb{Z}_2$, for simplicity.
The bulk magnetic fluxes lead to phenomenologically interesting aspects.
They lead to non-trivial profiles of bulk zero modes and can realize realistic hierarchical quark and lepton masses and their mixing angles.\footnote{See for quark and lepton mass matrices in magnetized orbifold models, e.g. Ref.~\cite{Hoshiya:2022qvr}.}
Thus, it is important to study the models with both bulk and localized fluxes.
In our future work, we would study the $T^2/\mathbb{Z}_2$ orbifold models with both 
bulk and localized fluxes such as flavor symmetries of both localized modes and bulk modes, 
the stability of the system, and realization of quark and lepton masses and their mixing angles.

\vspace{1.5 cm}
\noindent
{\large\bf Acknowledgments}\\

This work was supported by JPSP KAKENHI Grant Numbers JP23H04512 (H.O),  and JP23K03375 (T.K.), and JST SPRING Grant Number JPMJSP2119 (S.T.).


\appendix
\section{Summary of useful relations of wave functions}
\label{ap:relations}

First, we summarize the modular transformation of wave functions, $\psi^{(\alpha_1,\alpha_2)}(z,\tau)$~\cite{Kikuchi:2021ogn}.
Especially, they transform under $S$ and $T$ transformations as
\begin{align}
\begin{array}{cc}
\psi^{(0,0)}(S(z,\tau)) = (-\tau)^{1/2} e^{\pi i/4} \psi^{(0,0)}(z,\tau), & \psi^{(0,0)}(T(z,\tau)) = \psi^{(0,1/2)}(z,\tau), \\
\psi^{(1/2,0)}(S(z,\tau)) = (-\tau)^{1/2} e^{\pi i/4} \psi^{(0,1/2)}(z,\tau), & \psi^{(1/2,0)}(T(z,\tau)) = e^{\pi i/4} \psi^{(1/2,0)}(z,\tau), \\
\psi^{(0,1/2)}(S(z,\tau)) = (-\tau)^{1/2} e^{\pi i/4} \psi^{(1/2,0)}(z,\tau), & \psi^{(0,1/2)}(T(z,\tau)) = \psi^{(0,0)}(z,\tau), \\
\psi^{(1/2,1/2)}(S(z,\tau)) = (-\tau)^{1/2} e^{3\pi i/4} \psi^{(1/2,1/2)}(z,\tau), & \psi^{(1/2,1/2)}(T(z,\tau)) = e^{\pi i/4} \psi^{(1/2,1/2)}(z,\tau),
\end{array}
\label{eq:eachmodulartransf}
\end{align}

Next, we summarize the relations of wave functions, $\psi^{(\alpha_1,\alpha_2)}$, at $z$ and $Z_{z_{\rm f.p.} } = z - z_{\rm f.p.}$ as follows, 
\begin{align}
&
\begin{array}{l}
\psi^{(0,0)}(Z_{1/2},\tau) = e^{i \chi_1} \psi^{(0,1/2)}(z,\tau) \\
\psi^{(1/2,0)}(Z_{1/2},\tau) = e^{i \chi_1} e^{-\pi i/2} \psi^{(1/2,1/2)}(z,\tau) \\
\psi^{(0,1/2)}(Z_{1/2},\tau) = e^{i \chi_1} \psi^{(0,0)}(z,\tau) \\
\psi^{(1/2,1/2)}(Z_{1/2},\tau) = e^{i \chi_1} e^{-\pi i/2} \psi^{(1/2,0)}(z,\tau)
\end{array}, \label{eq:Z12} \\
&
\begin{array}{l}
\psi^{(0,0)}(Z_{\tau/2},\tau) = e^{i \chi_2} \psi^{(1/2,0)}(z,\tau) \\
\psi^{(1/2,0)}(Z_{\tau/2},\tau) = e^{i \chi_2} \psi^{(0,0)}(z,\tau) \\
\psi^{(0,1/2)}(Z_{\tau/2},\tau) = e^{i \chi_2} e^{-\pi i} \psi^{(1/2,1/2)}(z,\tau) \\
\psi^{(1/2,1/2)}(Z_{\tau/2},\tau) = e^{i \chi_2} \psi^{(0,1/2)}(z,\tau)
\end{array}, \label{eq:Ztau2} \\
&
\begin{array}{l}
\psi^{(0,0)}(Z_{(\tau+1)/2},\tau) = e^{i \chi_{1+2}} e^{-3\pi i/4} \psi^{(1/2,1/2)}(z,\tau) \\
\psi^{(1/2,0)}(Z_{(\tau+1)/2},\tau) = e^{i \chi_{1+2}} e^{-\pi i/4} \psi^{(0,1/2)}(z,\tau) \\
\psi^{(0,1/2)}(Z_{(\tau+1)/2},\tau) = e^{i \chi_{1+2}} e^{\pi i/4}\psi^{(1/2,0)}(z,\tau) \\
\psi^{(1/2,1/2)}(Z_{(\tau+1)/2},\tau) = e^{i \chi_{1+2}} e^{-\pi i/4} \psi^{(0,0)}(z,\tau)
\end{array}.
\label{eq:Ztau12}
\end{align}



\end{document}